\documentclass[%
 reprint,
 amsmath,amssymb,
 aps,
prb,
nofootinbib]{revtex4-2}

\usepackage{graphicx}
\usepackage{dcolumn}
\usepackage{bm}
\usepackage{diagbox}
\usepackage{braket}

\begin{document}

\preprint{APS/123-QED}

\title{Analytical Solution for the Steady States of the Driven Hubbard Model}
\author{J. Tindall$^{1}$, F. Schlawin$^{2,3}$, M. A. Sentef$^{3}$ and D. Jaksch$^{1}$}
\affiliation{$^1$Clarendon Laboratory, University of Oxford, Parks Road, Oxford OX1 3PU, United Kingdom} 
\email{joseph.tindall@physics.ox.ac.uk}
\affiliation{$^2$The Hamburg Centre for Ultrafast Imaging, Luruper Chaussee 149, D-22761 Hamburg, Germany}

\affiliation{$^3$Max Planck Institute for the Structure and Dynamics of Matter, Luruper Chaussee 149, D-22761 Hamburg, Germany}

\date{\today}

\begin{abstract}
Under the action of coherent periodic driving a generic quantum system will undergo Floquet heating and continously absorb energy until it reaches a featureless thermal state. The phase-space constraints induced by certain symmetries can, however, prevent this and allow the system to dynamically form robust steady states with off-diagonal long-range order. In this work, we take the Hubbard model on an arbitrary lattice with arbitrary filling and, by simultaneously diagonalising the two possible SU(2) symmetries of the system, we analytically construct the correlated steady states for different symmetry classes of driving. This construction allows us to make verifiable, quantitative predictions about the long-range particle-hole and spin-exchange correlations that these states can possess. In the case when both SU(2) symmetries are preserved in the thermodynamic limit we show how the driving can be used to form a unique condensate which simultaneously hosts particle-hole and spin-wave order.
\end{abstract}

\maketitle


\section{\label{sec:level1}Introduction}
Coherent driving has established itself as a fundamental tool for controlling and manipulating the states of quantum systems, from implementing high fidelity gates in few-qubit systems \cite{QubitGates} to inducing phase transitions in many-body optical lattices \cite{OpticalPhaseTransition}. Within this paradigm, recent experiments have observed how intense laser pulses in the midinfrared regime can transiently induce superconducting features - such as the opening of a gap in the real part of the optical conductivity and vanishing resistivity - when driving various solid state materials out of equilibrium \cite{OpticalSCExperiment1, OpticalSCExperiment2, OpticalSCExperiment3, OpticalSCExperiment4, OpticalSCExperiment5, OpticalSCExperiment6, OpticalSCExperiment7, OpticalSCExperiment8, KappaSaltExperiment}.
\par In order to understand seminal results such as these - and more generally the role coherent driving plays in altering the microscopic properties of many-body systems - significant theoretical studies have been undertaken. Floquet theory can be used to understand how periodic driving can modify the parameters of the system and create additional terms on top of the undriven Hamiltonian. This renormalization results in an effective Hamiltonian which on transient scales can, for example, favour superconducting prethermal states \cite{Coulthard, SCTheory1, SCTheory2, SCTheory3,SCTheory4,SCTheory5,SCTheory6, SCTheory7, SCTheory8, ScTheory9, SCTheory10, SCTheory11, SCTheory12, SCTheory13}, suppress wavepacket spreading and induce dynamical localization in a many-body bosonic gas \cite{dynamicallocalisationmanybody}, control spin-charge separation in a fermionic system \cite{Hongmin} or stabilise exotic spin-liquid states in frustrated systems \cite{Diehl}.  
\par It is inevitable, however, that due to Floquet heating a generic periodically driven quantum system will continuously absorb energy from the driving field. This heating competes with any transient order established by the effective Hamiltonian, causing it to melt away and leading to the formation of a featureless, infinite temperature state in the long-time limit \cite{LongTimeDriving1, LongTimeDriving2}. As a result, engineering Floquet Hamiltonians which are stable to heating on long timescales, allowing their prethermal states to be transiently observable is a current research endeavour attracting significant attention \cite{StablePrethermal1, StablePrethermal2, StablePrethermal3, Schiro1, StablePrethermal4, StablePrethermal5, StablePrethermal6, StablePrethermal7}.
\par In contrast to these efforts to mitigate the effects of heating in driven systems, recent theoretical work has shown how the presence of SU(2) symmetries in the fermionic Hubbard model can prevent featureless thermalisation and result in the formation of correlated, ordered states as the system heats up \cite{Buca2019, HeatingInducedOrder} - a mechanism termed `heating-induced order'. The phase space constraints induced by these symmetries mean the system is forced to relax towards steady states with off-diagonal long-range order as it absorbs energy from an external source. Currently, however, this has only been demonstrated numerically for the case when a single SU(2) symmetry is preserved in small, finite-size instances of the half-filled Hubbard chain \cite{HeatingInducedOrder}.
\par In this work we go beyond this, taking the driven Hubbard model on an arbitrary graph at arbitrary filling and analytically constructing the correlated steady states. We achieve this construction by simultaneously diagonalising the irreducible representation of the two possible SU(2) symmetries of the system and through it we can analytically calculate the steady state spin-exchange and particle-hole correlations, at any distance. We verify our analytical results with exact diagonalisation calculations and analyse how the long-time correlations depend on factors such as the filling, graph size, initial state and, crucially, the symmetries the driving possesses. Moreover, we provide the necessary conditions for the steady state correlations to remain finite in the thermodynamic limit. This leads us to show how, in cases where both SU(2) symmetries are preserved, the driving can be used to merge two independent condensates and create a unique spin-$\eta$ condensate which hosts both spin-exchange and particle-hole off-diagonal long-range order. Finally, we discuss possible experimental setups of the driven Hubbard model where the requisite symmetries are preserved in order to observe the formation of such unique, correlated states.

\section{\label{sec:level2}Theory}
As a starting point we consider the long-time states of a quantum system, with Hamiltonian $H$, subject to continued periodic driving under the modified Hamiltonian $H + H_{D}(t)$, where $H_{D}(t)$ is the time-dependent periodic driving term $H_{D}(t + T) = H_{D}(t)$. We then assume that we can find a set of $X$ operators $C = \{C_{1}, C_{2}, ..., C_{X}\}$, which form a linearly independent, irreducible representation of the symmetries of the system satisfying $[C_{i}, C_{j}] = 0 \ \forall i,j$. We note that whilst this is not always possible in a general quantum system, for the cases we consider in this work we are able to identify the completely commuting set $C$. Being a representation of the symmetries of the system we also clearly have that each member of $C$ satisfies 
\begin{equation}
[H + H_{D}(t), C_{i}] = 0, \ \ i = 1, ..., X.
\end{equation}
\par Under the action of $H + H_{D}(t)$ the system will continuously absorb energy until, in the long-time limit, it reaches a state of maximum entropy \cite{LongTimeDriving1}. The system is, however, under the constraint that its probability distribution over the eigenspace of the operators in $C$ must always be conserved. Consequently, the long-time state of the system will effectively have the form\footnote{Whilst clearly in a closed system a pure state will always remain pure, this mixed state ansatz is reasonable in a many-body system as the energy of the driving will have scrambled the phases of the wavefunction sufficiently to destroy any coherences between the $\ket{\alpha, \beta}$ eigenstates \cite{LongTimeDriving2}.}
\begin{align}
\lim_{t \rightarrow \infty}\rho(t)= \rho_{\infty} = \sum_{\alpha = (\alpha_{1}, \alpha_{2}, ... \alpha_{N})}P_{\alpha}\sum_{\beta = 1}^{D_{\alpha}}\ket{\alpha, \beta}\bra{\alpha, \beta},
\label{Eq:LongTimeState}
\end{align}
with $\sum_{\alpha}P_{\alpha}D_{\alpha} = 1$ and the multi-index/quantum number $\alpha$ running over the combinations of possible eigenvalues of the operators in $C$. For a given $\alpha$, the $D_{\alpha}$ vectors $\ket{\alpha, \beta}$ form the basis which simultaneously diagonalises the operators $C_{1}, C_{2}, ..., C_{X}$ and $P_{\alpha}$ is the probability of finding the state in this subspace. This probability must be preserved throughout the dynamics and thus
\begin{equation}
P_{\alpha}  = \sum_{\beta =1}^{D_{\alpha}}{\rm Tr}(\rho(0)\ket{\alpha, \beta}\bra{\alpha,\beta}),
\end{equation}
where $\rho(0)$ is the initial state of the system. If the complete basis $\{\ket{\alpha, \beta}\}$ can be constructed and the corresponding probabilities $P_{\alpha}$ calculated then the long-time state of the system is known. 
\par The structure and properties of the set $C$ has a significant influence on the properties of the steady state $\rho_{\infty}$. For example, consider the case of a many-body lattice with a single U(1) symmetry such as the total particle number. The corresponding total number operator is diagonal in the Fock basis and thus, by Eq. (\ref{Eq:LongTimeState}), so is $\rho_{\infty}$. All the states in the Fock basis can be written as a product state over the different lattice sites, making $\rho_{\infty}$ a featureless (outside of the well-defined particle number), unentangled thermal state. 
\par The same cannot be said, howevever, for more complicated symmetries. In the case of an SU(n) symmetry for a many-body lattice, one can form the set $C$ by using the generators to construct the $n-1$ Casimir operators which are fully independent of the lattice structure and thus completely translationally symmetric. The basis which simultaneously diagonalises these Casimir operators is the basis in which $\rho_{\infty}$ is diagonal. These basis states cannot all be written as product states over the different lattice sites and often contain correlations in the form of excitations which are spread between sites. Through Eq. (\ref{Eq:LongTimeState}) the long-time state will inherit these properties, along with the complete translational invariance of the Casimir operators, and thus possess correlations or excitations which are independent of the lattice geometry - i.e. they are completely uniform with distance. This induction of uniform, long-range correlations by heating an SU(n) symmetric system up to its steady state has been termed `heating-induced order' and has been studied numerically on small, finite-sized half-filled Hubbard lattices with a single preserved SU(2) symmetry \cite{Buca2019, HeatingInducedOrder}.
\par We note that the formation of these off-diagonal long-range ordered states as $t \rightarrow \infty$ does not exclude the possibility that, due to the driving, one could also observe the emergence of some transient, dynamical order on an intermediate timescale. This would then be followed by the melting of this order and the onset of our robust, steady state off-diagonal long-range order (ODLRO) once sufficient heating has occurred. The formation of transient, dynamical order is, however, usually reliant on a careful of choice of driving terms and parameters \cite{SCTheory2, SCTheory4,SCTheory6, SCTheory7, SCTheory10, SCTheory11, SCTheory12}. Meanwhile, if the requisite symmetries are satisfied, the emergence of steady state ODLRO is guaranteed due to the inevitability of Floquet heating under periodic driving - giving us significant freedom in the driving fields which can be used to observe heating-induced order.

\subsection{The Hubbard model on an arbitrary graph} 
\begin{figure}[t]
\centering
\includegraphics[width = \columnwidth]{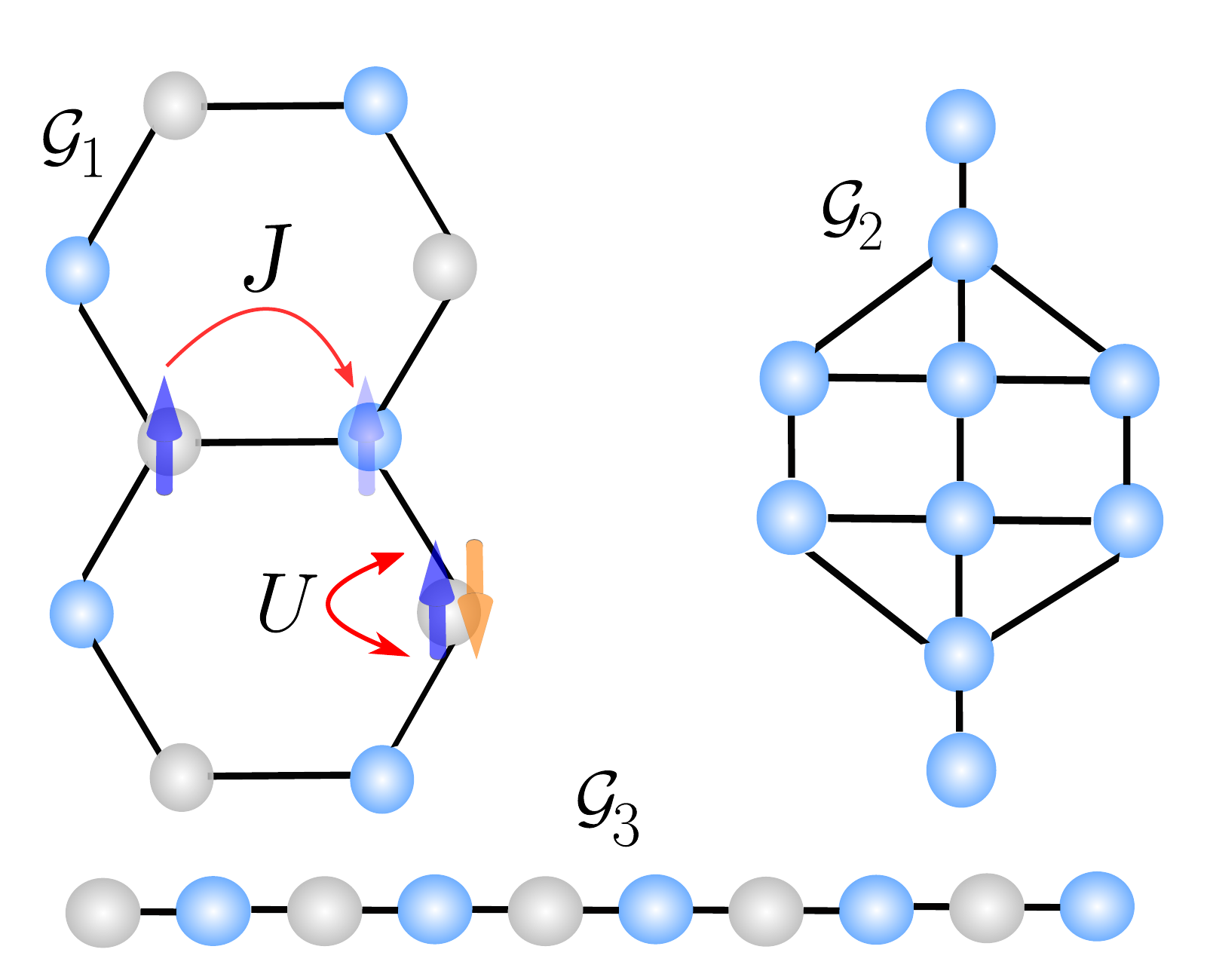}
\caption{Fermionic Hubbard model on a graph $\mathcal{G} = \mathcal{G}(V, E)$ where $V$ are the vertices and $E$ are the edges. The $M$ vertices form the lattice sites on which the fermions reside and interact with strength $U$ whilst the edges form the nearest neighbour bonds over which the fermions can hop with constant anplitude $\tau$. The Hamiltonian is defined in Eq. (\ref{Eq:ArbHubbard}) and $\mathcal{G}_{1}, \mathcal{G}_{2}$ and $\mathcal{G}_{3}$ are the $3$ different $M=10$ vertex graphs we use for our numerics. The grey vs blue sites represent a bi-partite splitting on the graphs $\mathcal{G}_{1}$ and $\mathcal{G}_{3}$.}
\label{Fig:F1}
\end{figure}

Here, we consider the Hubbard model on an arbitrary graph with arbitrary filling and, by simultaneously diagonalising the dual SU(2) symmetries, analytically construct the long-time states for the different symmetry classes of driving. We define this graph as $\mathcal{G} = (V,E)$ where $V$ are the vertices (or sites) and $E$ the edges. The Hamitonian can then be written as
\begin{equation}
H = -\tau\sum_{V, V'}\sum_{\sigma  = \uparrow, \downarrow}(c^{\dagger}_{\sigma, V}c_{\sigma, V'} + {\rm h.c}) + U\sum_{V}n_{\uparrow, V}n_{\downarrow, V},
\label{Eq:ArbHubbard}
\end{equation}
where $c_{\sigma, V}^{\dagger}$ and its adjoint are the usual creation and annihilation operators for a fermion of spin $\sigma$ on vertex $V$. The first summation runs over all the edges in the graph, kinetically coupling together the two sites $V, V'$ residing at the ends of each edge with strength $\tau$. The second summation runs over all the vertices in the graph and creates an energy penalty $U$ for vertices simultaneously occupied by both spin species. Additionally, $n_{\sigma, V}$ is the number operator for a particle of spin $\sigma$ on site $V$ and we use the integer $M$ to denote the number of vertices on the graph. We also depict this Hamiltonian in Fig. \ref{Fig:F1}, showing the 3 finite-size graphs we use for our numerical results, which serve to benchmark our analytical predictions. These analytical predictions, however, can be immediately applied to any graph of any number of vertices.

\par The Hamiltonian in Eq. (\ref{Eq:ArbHubbard}) has a rich symmetry structure comprised of either one or two ${\rm SU(2)}$ symmetries \cite{HubbardSymmetries}, which are fundamental to our results. The first, permanent, ${\rm SU}(2)$ symmetry can be introduced through the spin-raising operator $S^{+} = \sum_{V}c_{V, \uparrow}^{\dagger}c_{V, \downarrow}$, its conjugate $S^{-}$ and the total magnetisation $S^{z} = \sum_{V}n_{\uparrow, V} - n_{\downarrow, V}$. These operators are conserved over all graphs, i.e. $[H, S^{\pm, z}] = 0, \quad \forall \mathcal{G}$ and we refer to this symmetry as the `spin' symmetry with the corresponding operators only acting non-trivially on the sites of the lattice occupied by a single fermion (singlons). 

\par The second of the ${\rm SU}(2)$ symmetries is introduced through the $\eta$-raising operator $\eta^{+} = \sum_{V}f(V)c_{V, \uparrow}^{\dagger}c_{V, \downarrow}^{\dagger}$, its conjugate  $\eta^{-}$ and the modified total number operator $\eta^{z} = \sum_{V}(n_{\uparrow, V} + n_{\downarrow, V} - 1)$. If the graph $\mathcal{G}$ is bi-partite, i.e. the vertices can be split into two sets with the edges of the graph only forming connections between the two sets, then $[H, \eta^{+}\eta^{-}] = 0$ if we set the function $f(V)$ to take the values $\pm 1$ depending on whether the corresponding vertex is in the first or second set. Moreover, we always have $[H, \eta^{z}] = 0 \ \forall \mathcal{G}$ and thus the $\eta$-operators form an ${\rm SU}(2)$ or ${\rm U(1)}$ symmetry depending on whether the graph is bi-partite or not. The corresponding operators only act non-trivially on the empty and full sites within the lattice and these sites are often referred to as occupied by a `hole' or a `doublon' quasiparticle respectively.
\par In addition to this we note that there may be additional symmetries that $H$ posseses which will reflect the \textit{polygon} symmetries of the graph $\mathcal{G}$. For example on an open boundary $1$D chain the Hubbard model has a reflection symmetry about the central site, whilst on more complex geometries there may be mutiple reflection and translational shift symmetries \cite{HubbardSymmetries}.

Our goal is to determine the long-time states reached under continued periodic driving on top of the Hamiltonian in Eq. (\ref{Eq:ArbHubbard}). For simplicity, we will assume that the driving either breaks any \textit{polygon} symmetries in the system, or that they can be ignored due to their sufficiently small effect on the long-time properties of the system. We will see this is a reasonable assumption for the graphs we use and therefore the relevant symmetries to consider in our system are the two possible ${\rm SU}(2)$ symmetries of the lattice. 
\par From here on we will also, without loss of generality, fix the quantities $N_{\uparrow}$ and $N_{\downarrow}$ which correspond to the total number of fermions of spin $\uparrow$ and spin $\downarrow$ respectively. Our results in the full Hilbert space can be recovered by performing a direct sum over all possible values of $N_{\uparrow}$ and $N_{\downarrow}$. To keep the equations we derive more concise, we will restrict ourselves to lattices with an even number of sites $M$ and an even total number of particles $N = N_{\uparrow} + N_{\downarrow}$. For brevity we will also introduce the integer quantities $\alpha = (M-N)/2$ and $\beta = (N_{\uparrow} - N_{\downarrow})/2$

\subsection{Simultaneously diagonalising the Hubbard SU(2) Casimir operators} In order to construct the long-time state for arbitrary driving we must be able to diagonalise the irreducible representation of the dual SU(2) symmetries. This representation, i.e. our set $C$, consists of the two SU(2) Casimir operators $\eta^{2} = \eta^{+}\eta^{-} + \eta^{-}\eta^{+} + (\eta^{z})^{2}$ and $S^{2} = S^{+}S^{-} + S^{-}S^{+} + (S^{z})^{2}$ and their corresponding `z' operators $\eta^{z}$ and $S^{z}$. By fixing the particle numbers we have already removed the dependency on $\eta^{z}$ and $S^{z}$ and this restriction also means that the eigenvectors of the operator $O^{+}O^{-}$ are also those of $O^{-}O^{+}$, where $O$ is either $\eta$ or $S$. Hence, our problem is immediately simplified to simultaneously diagonalising $S^{+}S^{-}$ and $\eta^{+}\eta^{-}$. 
\par We can make progress with this problem by noticing that both operators commute with the total doublon operator $N_{D} = \sum_{V}n_{\uparrow}n_{\downarrow}$. We can therefore simultaneously reduce them into block matrices with the blocks indexed by $i$, the number of doublons on the graph, which ranges from ${\rm Max}(0, -2\alpha)$ to ${\rm Min}(N_{\uparrow}, N_{\downarrow})$. For a given value of $i$, we observe that there must be $M + i - N$ holes in the graph and so the remaining vertices, or sites, will be occupied by $N_{\uparrow} - i$ and $N_{\downarrow} - i$ singlons of spin $\uparrow$ and $\downarrow$ respectively. 
\par We can use this knowledge to take any given block and arrange the sites of the lattice into two sets, with the first set ($A$) containing $M + 2i - N$ sites and the second set ($B$) containing the remaining $N - 2i$ sites. There are ${M \choose N - 2i}$ different ways in which the sites can be arranged in this manner and in $A$ we place all of the doublons and holes whilst in $B$ we place all of the singlons. If the operator $S^{+}_{V}S^{-}_{V'}$ ($\eta^{+}_{V}\eta^{-}_{V'}$) acts on a vertex which is in set $A$ ($B$) then it will immediately annihilate any given basis state, and thus we can let
\begin{align}
 &\eta^{+}\eta^{-} \rightarrow (\eta^{+}\eta^{-})' = \sum_{V, V' \in A}\eta^{+}_{V}\eta^{-}_{V'}, \notag \\ &S^{+}S^{-} \rightarrow (S^{+}S^{-})' = \sum_{V, V' \in B}S^{+}_{V}S^{-}_{V'},
\end{align}
and ignore the other terms in these two operators. If we can now construct a state $\ket{\eta}$, within $A$, which is an eigenvector of $(\eta^{+}\eta^{-})'$ with eigenvalue $\lambda_{A}$ and a state $\ket{S}$ within $B$ which is an eigenvector of $(S^{+}S^{-})'$ with eigenvalue $\lambda_{B}$ then their `tensor product' $\ket{\eta} \tilde{\otimes} \ket{S}$ will simultaneously be an eigenvector of both $\eta^{+}\eta^{-}$ and $S^{+}S^{-}$ on the full graph; with eigenvalues $\lambda_{A}$ and $\lambda_{B}$ respectively. The tilde on the tensor product means we will take into account the way the sets $A$ and $B$ were formed and re-order the vertices of the graph back to their original order.

\par We now need to determine the eigenspectrum of $(\eta^{+}\eta^{-})'$ and $(S^{+}S^{-})'$. Crucially, we know that $(\eta^{+}\eta^{-})'$ and $(S^{+}S^{-})'$ correspond to the SU(2) Casimir operators in a, respectively, $M+2i-N$ and $N-2i$-fold tensor product representation of the fundamental representation of SU(2) \cite{SU2Ref1, SU2Ref2}. By exploiting the `ladder' structure of SU(2) representations we can determine the eigenvalues of $(\eta^{+}\eta^{-})'$ and $(S^{+}S^{-})'$ as
\begin{align}
&\lambda_{\eta}(k) = k(k+1) - \alpha (\alpha + 1), \quad k = |\alpha|, ... \alpha + i \notag \\
&\lambda_{s}(m) = m(m + 1) - \beta (\beta + 1), \quad m = |\beta|, ..., N/2 - i.
\end{align}
These eigenvalues have the following degeneracies
\begin{align}
&D_{\eta}(k) = C(\alpha + i + k, \alpha + i - k - \delta_{k}), \notag \\
&D_{S}(m) = C(N/2 -i + m, N/2 -i -m - \delta_{m}),
\end{align}

where $\delta_{a}$ is the Kronecker delta function and $C(x, y) \equiv C^{x}_{y}$ is the Catalan Triangle Number \cite{CatalanNumbers}
\begin{align}
C^{x}_{y} = \begin{cases}
\frac{(x+y)!(x-y+1)}{y!(x+1)!} &  x,y > 0, \\
\hfil 1 & \text{Otherwise.}
\end{cases}
\end{align}
With this knowledge we can denote $\ket{\eta_{k, l}}$ and $\ket{S_{m, n}}$ as the eigenvectors of $(\eta^{+}\eta^{-})'$ and $(S^{+}S^{-})'$ where $k$ and $m$ index the respective eigenvalues and $l$ and $n$ run through the degenerate eigenvectors for a given $k$ and $m$. The state $\ket{\psi} = \ket{\eta_{k, l}} \tilde{\otimes} \ket{S_{m, n}}$ is then an eigenvector of $\eta^{+}\eta^{-}$ and $S^{+}S^{-}$ on the full graph $\mathcal{G}$. The relevant indices to specify the state $\ket{\psi}$ are then $\ket{\psi_{i, j, k, l, m, n}}$ where $i$ is the number of doublons, $j$ indexes the ${L \choose N - 2i}$ ways in which the lattice can be split into the two aforementioned sets, $k$ and $l$ index the degenerate eigenvectors of $(\eta^{+}\eta^{-})'$ whilst $m$ and $n$ do the same for $(S^{+}S^{-})'$. The states $\ket{\psi}$ form a complete, orthonormal basis which diagonalise $\eta^{+}\eta^{-}$ and $S^{+}S^{-}$ for the given filling.
\newline 

\begin{table*}[t]
\begin{center}
\centering
\begin{tabular}{|l|*{2}{c|}}\hline
\backslashbox{\rm Spin Symmetry}{\rm $\eta$ \ Symmetry} & {\rm No} & {\rm Yes} \\
\hline \hfil {\rm No} & $P_{m, k} = {\rm const.}$ & $P_{m,k} = P_{m}P_{k}, \  P_{k} = {\rm Tr}(\rho(0) \mathcal{P}^{\eta}_{k}) \ P_{m} = {\rm const.}$ \\
\hline \hfil {\rm Yes} & $P_{m,k} = P_{m}P_{k}, \ P_{k} = {\rm const.}, \ P_{m} = {\rm Tr}(\rho(0) \mathcal{P}^{S}_{m}) $ & $P_{m,k} = {\rm Tr}(\rho(0) \mathcal{P}^{S, \eta}_{m, k})$ \\ \hline
\end{tabular}
\end{center}
\caption{Form of the probabilities $P_{k,m}$ which characterise the steady state of the driven Hubbard model (see Eq. (\ref{Eq:HubbardLongTimeState})). The probabilities are defined by their relationship to the initial state $\rho(0)$ and the projectors $\mathcal{P}^{\eta}_{k}$, $\mathcal{P}^{S}_{m}$ and $\mathcal{P}^{S, \eta}_{m, k}$, which are defined in Eq. (\ref{Eq:ProjectorsHubb}). This relationship changes depending on which of the two SU(2) symmetries are present during the systems evolution to the steady state.}
\label{Table:T1}
\end{table*}

\subsection{Steady states of the driven Hubbard model}
We can now combine the basis we have constructed with the general result of Eq. (\ref{Eq:LongTimeState}) and write down the steady state $\rho_{\infty}$ of the periodically driven Hubbard model on an arbitrary graph with arbitrary filling as 

\begin{widetext}
\begin{align}
\rho_{\infty} = \frac{1}{Z}\sum_{i = {\rm Max}(0, -2\alpha)}^{{\rm Min}(N_{\uparrow}, N_{\downarrow} )}\ \sum_{k=|\alpha|}^{\alpha + i}\ \sum_{m=|\beta|}^{N/2 -i}P_{k,m}\sum_{l=1}^{C^{\alpha + i+k}_{\alpha + i-k-\delta_{k}}} \sum_{n=1}^{C^{N/2-i+m}_{N/2-i-m-\delta_{m}}} \sum_{j = 1}^{L \choose N -  2i} \ket{\psi_{i, j, k, l, m, n}} \bra{\psi_{i, j, k, l, m, n}},
\label{Eq:HubbardLongTimeState}
\end{align}
\end{widetext}
where $Z$ is the partition function and the values $P_{k, m}$ are a series of probabilities (analogous to the $P_{\alpha}$ in Eq. (\ref{Eq:LongTimeState})) which are are dependent on both the initial state as well as which, if any, of the SU(2) symmetries are present in the system. This dependency is encapsulated by Table \ref{Table:T1}, where we have also introduced the following projectors
\begin{align}
&\mathcal{P}^{\eta}_{k} = \sum_{i, j, m, n, l}\ket{\psi_{i, j, k, l, m, n}} \bra{\psi_{i, j, k, l, m, n}}, \notag \\
&\mathcal{P}^{S}_{m} = \sum_{i, j, k, n ,l}\ket{\psi_{i, j, k, l, m, n}} \bra{\psi_{i, j, k, l, m, n}}, \notag \\
&\mathcal{P}^{S, \eta}_{m, k} = \sum_{i,j,n,l}\ket{\psi_{i, j, k, l, m, n}} \bra{\psi_{i, j, k, l, m, n}},
\label{Eq:ProjectorsHubb}
\end{align}
with the indices used all retaining their original meaning and ranges from Eq. (\ref{Eq:HubbardLongTimeState}).

\par Table \ref{Table:T1}, alongside Eq. (\ref{Eq:HubbardLongTimeState}), allows us to classify and write down the steady states of the driven Hubbard model on an arbitrary graph. This classification is based on which of the SU(2) symmetries are present during the time evolution and - due to the excited, entangled nature of the eigenstates of the SU(2) Casimir operators - the long-time state will contain long-range correlations in the channels corresponding to the preserved symmetries. Meanwhile, in the channels where the underlying symmetry was not present the constant nature of the probabilities means all excitations are equally likely and they will destructively interfere with each other to ensure there is no long-range order in that channel.
\par Given the probabilities $P_{k, m}$ we can calculate a number of properties of the state in Eq. (\ref{Eq:HubbardLongTimeState}). For example we can immediately deduce the moments of the doublon number $N_{d}$
\begin{align}
&\langle N_{d}^{\alpha} \rangle= \frac{1}{Z}\sum_{i}i^{\alpha}f(i), \notag \\
&f(i) = \sum_{k, m}P_{k,m}C^{\alpha + i+k}_{\alpha + i-k-\delta_{k}}C^{N/2-i+m}_{N/2-i-m-\delta_{m}}{M \choose N -  2i}.
\label{Eq:HubbardLongTimeStateMoments}
\end{align}
These equations are useful because we can take advantage of the distance-invariance of correlations in the long-time state and use the first moment of the doublon number ($\alpha = 1$), along with the initial values $\langle \eta^{+}\eta^{-} \rangle$ and $\langle S^{+}S^{-} \rangle$, to directly extract values for the off-diagonal spin-exchange and particle-hole order parameters $\langle S^{+}_{V}S^{-}_{V'} \rangle$ and $\langle \eta^{+}_{V}\eta^{-}_{V'} \rangle$. Specifically, we know that
\begin{equation}
    \langle \eta^{+}\eta^{-} \rangle = \langle N_{d}^{1} \rangle + M(M-1)\langle \eta^{+}_{V}\eta^{-}_{V'} \rangle,
    \label{Eq: Correlations}
\end{equation}
and 
\begin{equation}
    \langle S^{+}S^{-} \rangle = N_{\uparrow} -\langle N_{d}^{1} \rangle + M(M-1)\langle S^{+}_{V}S^{-}_{V'} \rangle,
    \label{Eq: Correlations2}
\end{equation}
where $V \neq V'$. Moreover, higher moments of the doublon number, $\alpha > 1$ provide access to multi-point correlators in the $\eta$ and spin symmetry sectors (for example we can show that $\langle n_{\uparrow, V}n_{\downarrow, V}n_{\uparrow, V'}n_{\downarrow, V'} \rangle \propto \langle N_{d}^{2} \rangle - \langle N_{d}^{1} \rangle$).
\par In principle, however, in order to calculate the moments of the doublon number we need to know the exact values of the $P_{k,m}$s - which, in some cases, could be quite complicated and would involve taking a number of projective measurements on the initial state. We find from our equations, however, that the first moment of the doublon number is only dependent on the probabilities $P_{k,m}$ through its relationship to $\langle \eta^{+}\eta^{-}\rangle$ and $\langle S^{+}S^{-}\rangle$ and thus knowledge of these two values, and the graph size and filling, is enough to calculate $\langle N_{d}^{1} \rangle$. The steady state off-diagonal order parameters  $\langle \eta^{+}_{V}\eta^{-}_{V'} \rangle$ and $\langle S^{+}_{V}S^{-}_{V'} \rangle$ then follow immediately from Eq. (\ref{Eq: Correlations}) and the corresponding `spin' version.
\par These quantities are particularly important because when finite in the thermodynamic limit and completely uniform with distance, the latter of which is automatically satisfied by $\rho_{\infty}$, they describe the existence of a spin-wave or $\eta$ condensate. These condensates are underpinned by excitations which are completely spread out in space and in an $\eta$ condensate the long-range finite value of $\langle \eta^{+}_{V}\eta^{-}_{V'} \rangle$ directly implies superconductivity as the Meissner effect and flux quantisation can be observed \cite{MeissnerFlux1, MeissnerFlux2}. 

\section{Results - Hubbard Model under Generic Driving}
We now demonstrate the results we have derived explicitly, first focussing on the case where a single SU(2) symmetry is preserved and then moving on to the case where both SU(2) symmetries are preserved. We note that the scenario where both symmetries are not preserved needs no attention as it is trivial and the long-time state will simply be a featureless thermal state with a fixed particle number.

\begin{figure}[t]
\centering
\includegraphics[width = \columnwidth]{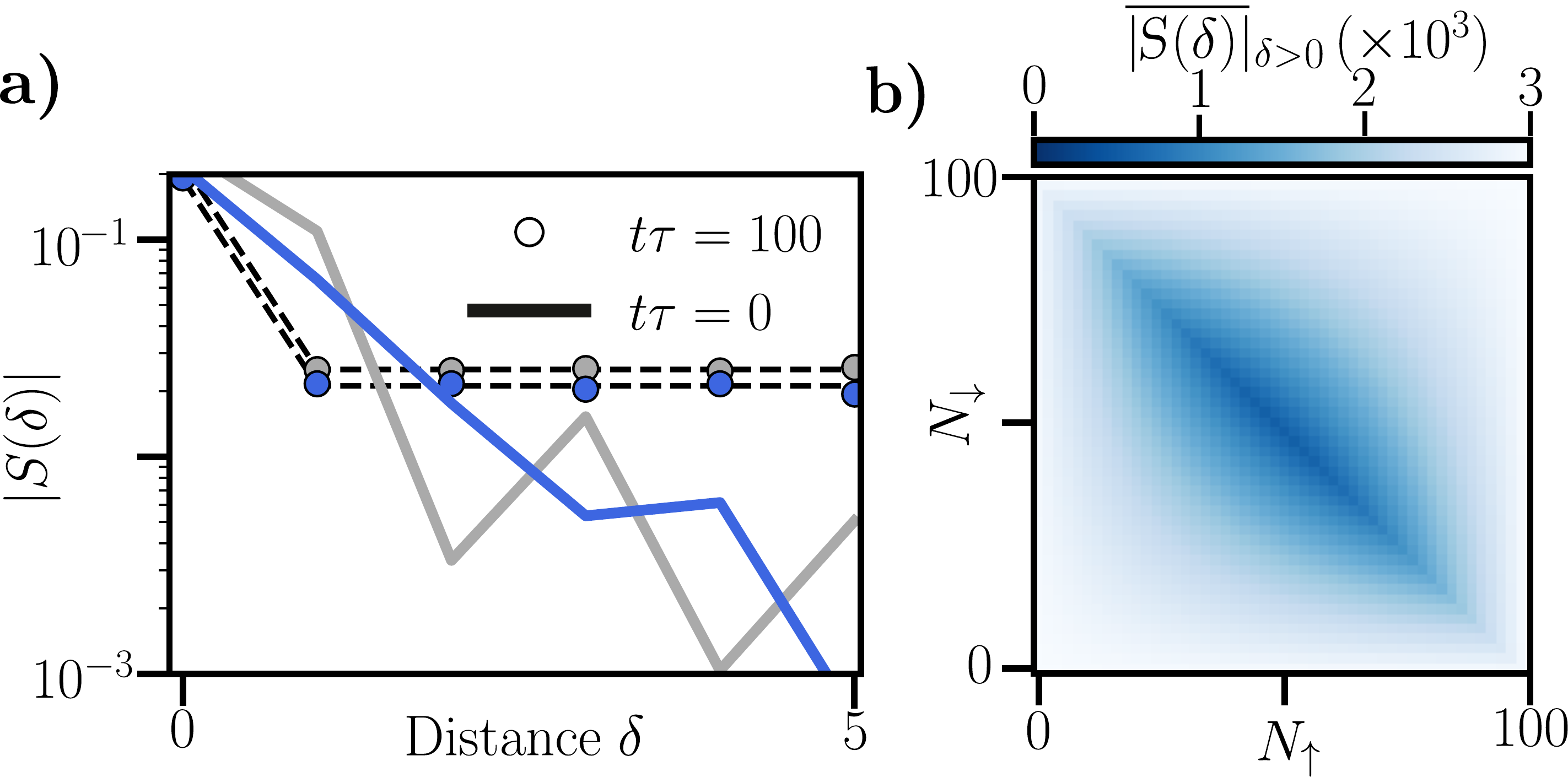}
\caption{(a) Spin-exchange correlations versus distance $\delta$, see Eq. (\ref{Eq:CorrelationFunction}), for the periodically driven graph $\mathcal{G}_{1}$ pictured in Fig. \ref{Fig:F1}, with the site index $V$ running from $1$ to $10$, starting at $V=1$ for the bottom left site and increasing in an anti-clockwise manner. The driving term we use is of the form $H_{D} = A\cos(\Omega t)\sum_{V}Vn_{V}$ and the grey vs blue markers/lines correspond to the fillings $N_{\uparrow} = N_{\downarrow} = 5$ and $N_{\uparrow} = N_{\downarrow} = 3$ respectively. At time $t\tau = 0$ the system is initialised in the  ground state of the undriven Hamiltionian $H$ with $U = 0.5\tau$ and then evolved under the Hamiltonian $H + H_{D}$ with $U = \tau$, $A = 6.0\tau$ and $\Omega = 1.0\tau$. The solid lines vs markers indicate the correlations at $t\tau = 0$ and $t\tau = 100$ respectively. The black-dotted lines give the exact results in the long-time limit. b) Off-diagonal spin-exchange order as a function of the numbers of $\uparrow$ and $\downarrow$ fermions in the steady state of a driven, arbitrary, $M = 100$ vertex Hubbard graph where the spin SU(2) symmetry is preserved whilst the $\eta$ SU(2) symmetry isn't. The system is initialised with $\langle S^{+}S^{-} \rangle = 0$.}
\label{Fig:F3}
\end{figure}

\subsection{Single Symmetry Preservation} We consider a setup in which the spin ${\rm SU}(2)$ symmetry is present whilst the $\eta$ ${\rm SU}(2)$ symmetry is not. These results are immediately analagous to the case where the spin-symmetry is not present and the $\eta$ symmetry is, which was studied in Ref. \citep{HeatingInducedOrder} but only for a small, half-filled chain. We introduce the graph measure $d(V, V')$ which is the minimum number of edges that must be traversed to move between the vertices $V$ and $V'$. With this measure we can then define the correlation function
\begin{equation}
    O(\delta) = \frac{1}{\mathcal{N}}\sum_{\substack{\langle V, V' \rangle \\ d(V, V') = \delta}}\langle O^{+}_{V}O^{-}_{V'}\rangle,
\label{Eq:CorrelationFunction}
\end{equation}
where $O$ is either $S$ or $\eta$, the summation is over all pairs of vertices where $d(V, V') = \delta$ and $\mathcal{N}$ is the number of pairs of vertices which satisfy $d(V, V') = \delta$. Hence, $O(\delta)$ measures the average of the spin-exchange or particle-hole correlations at a distance $\delta$ for any given graph and we can introduce $\overline{\vert O(\delta) \vert}_{\delta > l}$ as the average magnitude of these correlations at distances greater than $l$.
\par In Fig. \ref{Fig:F3}a we demonstrate agreement, for two different particle fillings, between the equations in the previous section and exact diagonalisation code which reaches the long-time limit of the graph $\mathcal{G}_{1}$ from Fig. \ref{Fig:F1}. As the driving explicitly breaks $\eta$ symmetry the long-time prediction from these equations is independent of the lattice structure and whether it is bi-partite or not. Figure \ref{Fig:F3} shows that when driving the ground state out of equilibrium, the preservation of $\langle S^{+}S^{-}\rangle$ under driving causes the establishment of completely uniform, long-range spin-wave order in the long-time limit, with a significant enhancement of the long-range correlations. This long-range order is largest at the higher filling and, more generally, our equations show that the steady state order will always be maximised when the system is closest to $0$ total magnetisation and half-filling, where the largest number of spin-exchange excitations are available. In Fig. \ref{Fig:F3}b we show this result explicitly, observing significant order around this half-filled non-magnetic point which then decays to $0$ as either of the filling numbers, $N_{\uparrow}$ and $N_{\downarrow}$, approach their maximal or minimal values (where it is not possible to perform a spin-exchange $S^{+}_{V}S^{-}_{V'}$ without annihilating the state of the system).

\begin{figure}[t]
\centering
\includegraphics[width = \columnwidth]{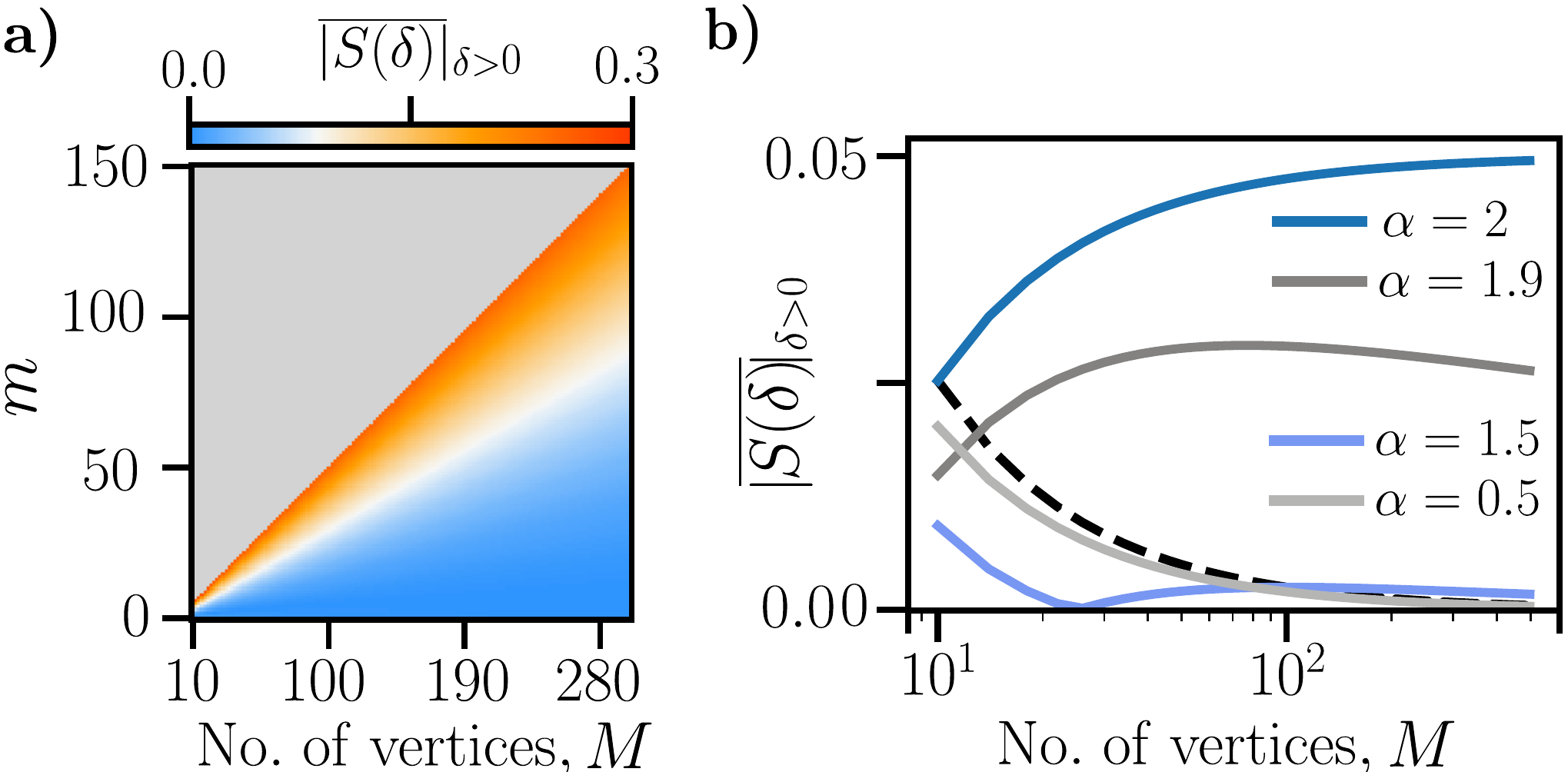}
\caption{Scaling of the steady state off-diagonal spin-exchange correlations with the number of vertices and initial value of $\langle S^{+}S^{-}\rangle = {\rm Tr}(\rho(0)S^{+}S^{-})$ for the half-filled Hubbard model on an arbitrary graph. The system is time evolved under driving which preserves the spin SU(2) symmetry and breaks the $\eta$ symmetry. a) Scaling versus both system size and $m$ where $m = (-1 + \sqrt{1 + 4\langle S^{+}S^{-}\rangle})/2$. b) Scaling with system size for various initial states. The solid lines correspond to the initial values $\langle S^{+}S^{-}\rangle = M^{\alpha}/20$ for varying $\alpha$ whilst the dashed line is for $\langle S^{+}S^{-}\rangle = 0$}
\label{Fig:F4}
\end{figure}

\begin{figure*}[t]
\centering
\includegraphics[width = \textwidth]{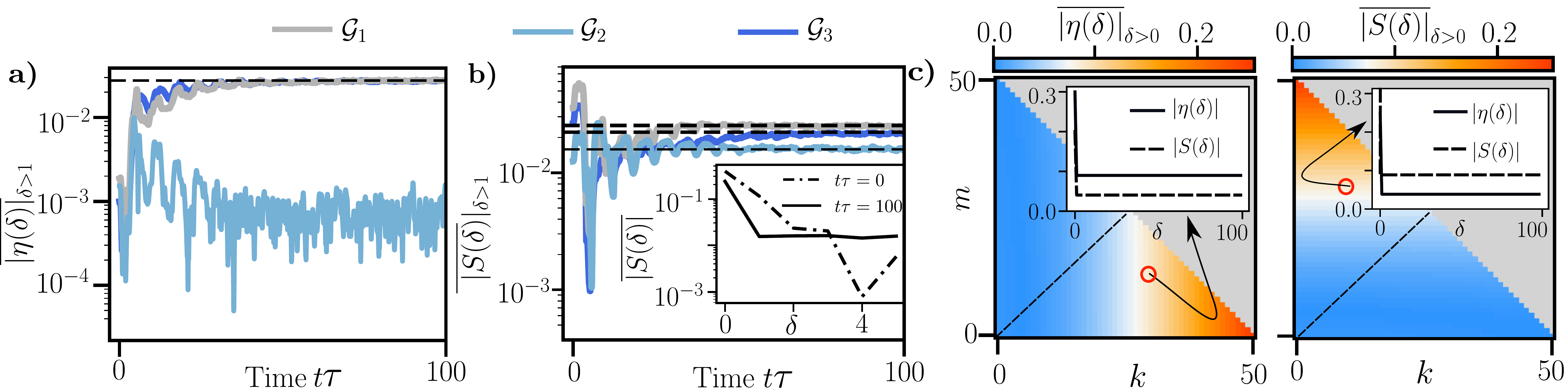}
\caption{a-b) Off-Diagonal Spin and $\eta$ Correlations versus time and distance for $3$ different half-filled Hubbard lattices with driving of the form $H_{D} = \delta U\cos(\omega t)\sum_{i}n_{\uparrow, i}n_{\downarrow, i}$ and the bare Hamiltonian $H$ as defined in Eq. (\ref{Eq:ArbHubbard}) with $U = 4.0\tau$. The system is initialised, at time $t\tau = 0$,  in the thermal state $\rho \propto \exp(-\beta H)$ with $\beta\tau = 5$ and then time-evolved under $H$ with $U = 4.0\tau$, $\delta U = 1.5\tau$ and $\omega = 1.0\tau$. Black-Dotted lines represent the long-time analytical predictions for the $3$ respective graphs. Inset) Spin-correlations versus distance at times $t\tau = 0$ and $t\tau = 100$ for the lattice $\mathcal{G}_{2}$. c) Map of the doublon/ spin order of long-time states of the Hubbard model on a $100$ vertex bi-partite graph with driving which preserves both SU(2) symmetries. The indices $m$ and $k$ are dependent on the initial values of the spin and $\eta$ symmetry via $m = (-1 + \sqrt{1 + 4\langle S^{+}S^{-}\rangle})/2$ and $k = (-1 + \sqrt{1 + 4\langle \eta^{+}\eta^{-}\rangle})/2$. The two maps are related via a reflection over the black-dotted line. Insets: Doublon (Solid Line) and Spin (Dashed Line) order for the long-time state at the circled point on the map.}
\label{Fig:F5}
\end{figure*}

\par We then concentrate on the non-magnetic half-filled point, where the order is maximised, and in Fig. \ref{Fig:F4} plot the magnitude of the long-time spin-exchange order for a large range of system sizes and initial values ${\rm Tr}(\rho(0)S^{+}S^{-})$. These results apply to \textit{any} driven Hubbard graph where the spin SU(2) symmetry is preserved and the $\eta$ symmetry isn't - which could be a result of the driving or the underlying lattice structure. 
\par Notably, as system size increases there is a growing space of initial values where there is significant spin-wave order in the long-time limit. In fact we can easily argue that the spin-wave order will be finite in the thermodynamic limit of any graph if, and only if, the initial state satisfies $\langle S^{+}S^{-}\rangle \propto M^{2}$, with Fig. \ref{Fig:F4}b showing this explicitly. Such states are already likely to have finite long-range spin correlations to satisfy this requirement, however the driving will still act to renormalize these correlations and make them completely uniform with distance, stabilizing the spin-wave order. Meanwhile, for simulations which start in the ground state (where $\langle S^{+}S^{-} \rangle \equiv 0$) the spin order asymptotically tends to $0$ as $1/M$ but remains finite for any finite-size system - with the dynamics underpinned by a drastic amplification of the long-range correlations at the expense of the short-range ones.

\par Interestingly, in Ref. \cite{Coulthard}, a type of spin preserving periodic driving was studied for the 1D Hubbard chain in the thermodynamic limit. There it was shown how driving the ground state can renormalize the exchange parameters in the system and transiently enhance long-range singlet pairing. Even though the long-time spin-order will be $0$, our results here suggest this transient response could be a result of the preservation of $\langle S^{+}S^{-} \rangle$. Under driving, this preservation forces a drastic reorganisation of the spin degrees of freedom, which will involve a transient enhancement of the long-range correlations at the expense of the shorter ones, before they mutually decay away to $0$ in the long-time limit.

\par Despite the ground state of hypercubic Hubbard lattices possessing the smallest possible value of $\langle S^{+}S^{-} \rangle$, our equations show that the magnitude of the induced off-diagonal spin order $\vert \langle S^{+}_{V}S^{-}_{V'} \rangle \vert $ under driving which preserves the SU(2) spin symmetry is larger than that for any initial states which have finite $\langle S^{+}S^{-} \rangle < (M-1)/2$. This is because the steady state spin order $\langle S^{+}_{V}S^{-}_{V'} \rangle$ is a monotonic function of $\langle S^{+}S^{-} \rangle$ but is negative for finite $M$ and $\langle S^{+}S^{-} \rangle < M/4$, at which point it changes sign. Hence, the ground state spin order is the most negative and it can be shown from our equations that the magnitude of this order is larger than that of any other states in the range $0 < \langle S^{+}S^{-} \rangle < (M-1)/2$. On a half-filled hypercubic lattice, this range includes all initial states in thermal equilibrium $\rho(0) \propto e^{-\beta H}$ as the value of $\langle S^{+}S^{-} \rangle$ for these states monotonically increases from $0$ to $M/4$ as the inverse temperature $\beta$ decreases from $\infty$ to $0$. Hence, for these thermal initial states on a finite sized lattice, the magnitude of the steady state spin order is maximised for the ground state and remains finite for any finite temperature initial state - asymptotically tending to $0$ as the temperature of the initial state approaches $\infty$ where $\langle S^{+}S^{-} \rangle = M/4$.

\par Outside of hypercubic lattices it is harder to make statements about the long-time order which will form from an initial state in thermal equilibrium. This is because the relationship between the expectation value of the SU(2) Casimir operator (which determines the amplitude of the long-time order) and $\beta$ is likely to be more complex. This could, however, lead to the exciting possibility of heating $\rho(0) \propto e^{-\beta H}$ under certain symmetries and forming a state with uniform, finite, off-diagonal order even in the thermodynamic limit – dynamically transforming a system in thermodynamic equilibrium into a `hot’ condensate.

\par It is worth emphasizing that in this paper we have taken the hopping strength $\tau$ to be homogeneous across all edges of the lattice. The spin SU(2) symmetry of the Hubbard model is, however, preserved even in the case of an inhomogeneous hopping strength and so our results in this section immediately apply to this more general scenario. 

\begin{figure*}[t]
\centering
\includegraphics[width = \textwidth]{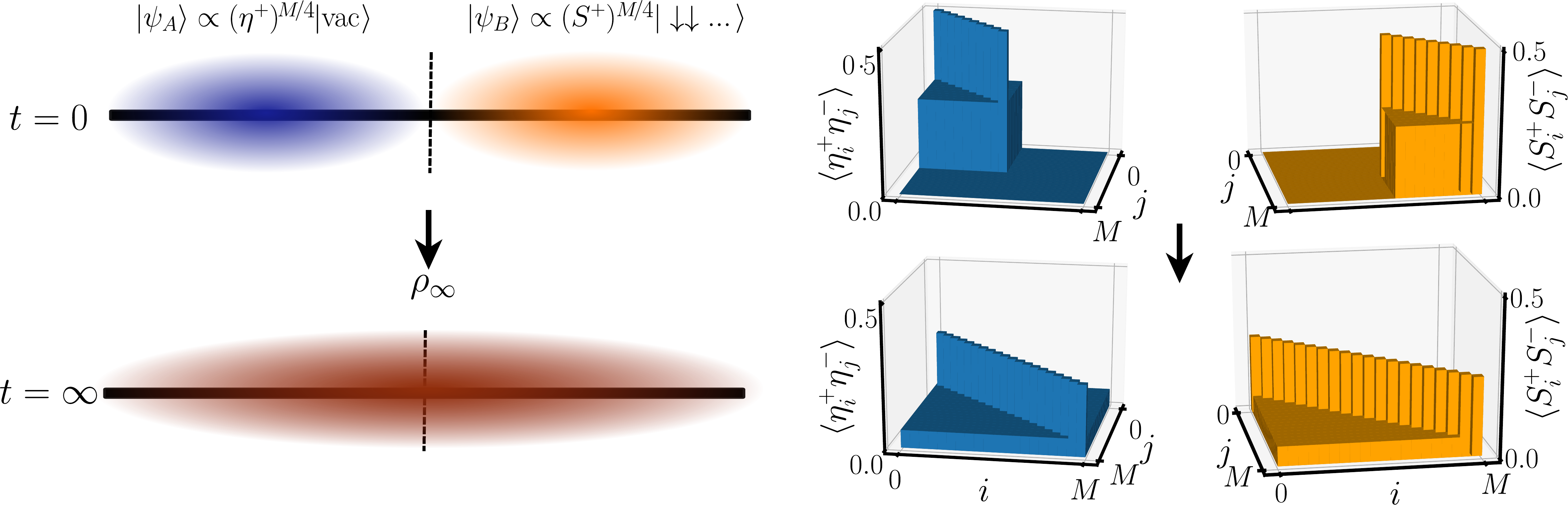}
\caption{Merging of two condensates under long-time driving. The Hubbard model on a 1D chain is split into two independent halves which contain an $\eta$ and spin-wave condensate respectively. The system is then time-evolved under generic driving which respects both SU(2) symmetries of the model, causing the condensates to merge into a single hybrid spin-$\eta$ condensate. Left) Pictorial depiction of the process. Right) Matrices of $\eta$ and spin- exchange correlations for the whole lattice in the thermodynamic limit ($M \rightarrow \infty$) at times $t=0$ and $t = \infty$, $i$ and $j$ index the different sites of the chain.}
\label{Fig:F6}
\end{figure*}

\subsection{Both SU(2) Symmetries Preserved} 
We now move to the case where the driving preserves both ${\rm SU}(2)$ symmetries and in order for this preservation to also be true for the full Hamiltonian $H + H_{D}(t)$ we require the underlying graph to be bi-partite, which we will assume in our analytical calculations. For our numerics, the graphs $\mathcal{G}_{1}$ and $\mathcal{G}_{3}$ are bi-partite whilst $\mathcal{G}_{2}$ isn't and so we shall see that this leads to distinctly different dynamics when they are driven. We fix ourselves to symmetric half-filling ($N_{\uparrow} = N_{\downarrow} = M/2$) where the long-time order will be maximal due to the maximum availability of both particle-hole and spin-exchange excitations. In Figure \ref{Fig:F5}a-b. we start in a thermal state and apply driving which preserves both SU(2) symmetries to the three different lattice structures $\mathcal{G}_{1}, \mathcal{G}_{2}$ and $\mathcal{G}_{3}$, calculating the average amplitude of the correlations between all pairs of sites not connected by an edge. In the particle-hole channel, we observe a pronounced increase in this average, with the large value of $U$ having suppressed them in the initial thermal state. The average in the spin-exchange sector instead remains relatively constant, but the inset in Fig. \ref{Fig:F5}c shows that the driving has still reordered these correlations to be completely uniform with distance. We find our predictions from Eq. (\ref{Eq:HubbardLongTimeStateMoments}) are in perfect quantitative agreement with the long-time order observed in these states and, as we expect, the non-bipartite graph $\mathcal{G}_{2}$ cannot support particle-hole order as it lacks the requisite symmetry.
\par In Fig. \ref{Fig:F5}c we present maps of the long-time off-diagonal particle-hole and spin-exchange order as a function of the $\eta$ and spin eigenvalues of the initial state for a bi-partite graph with $M = 100$ sites. There is a whole manifold of states with significant, co-existing, particle-hole and spin-exchange order, and we present an example state in the inset of each map. Here, we can also show that, similar to the single symmetry case, when starting in the ground state the off-diagonal spin and $\eta$ order always remains finite for finite systems but asymptotically decays to $0$ as $1/M$.

\par In fact, in the thermodynamic limit, the condition  $\langle \eta^{+}\eta^{-} \rangle \propto M^{2}$ or $\langle S^{+}S^{-} \rangle \propto M^{2}$ is necessary to observe finite ODLRO in the $\eta$ and spin sectors respectively. These two conditions are not mutually exclusive, allowing us to exploit the driving to form a unique spin-$\eta$ condensate. Specifically, consider a bi-partite $M$ site lattice which hosts the initial state $(S^{+})^{M/4}\ket{\chi_{1}} \otimes (\eta^{+})^{M/4}\ket{\chi_{2}}$, where $\ket{\chi_{1}} = \ket{\downarrow, \downarrow, ... , \downarrow}$ on $M/2$ of the sites and $\ket{\chi_{2}}$ is the vacuum state on the other $M/2$ sites. This state consists of two independent condensates - a spin-wave and particle-hole condensate - each confined to one half of the lattice. Under driving which preserves both SU(2) symmetries, and a Hubbard Hamiltonian $H$ over the full lattice, the dynamics will involve the system heating up whilst conserving the values of $\langle S^{+}S^{-} \rangle$ and $\langle \eta^{+}\eta^{-}\rangle$ - forcing the condensates to merge and phase-lock into a larger, single condensate which, remarkably, hosts ODLRO in the particle-hole and spin-exchange sectors simultaneously. Specifically, our equations tell us $\lim_{M \rightarrow \infty}\overline{\vert \eta(\delta) \vert}_{\delta > 0} = \lim_{M \rightarrow \infty}\overline{\vert S(\delta) \vert}_{\delta > 0} = 0.0625$, i.e. there is completely ODLRO in both symmetry sectors. The interplay between heating induced by the driving and the preservation of SU(2) symmetries has therefore led to the formation of a unique state of matter which is simultaneously a superconductor and a spin-wave condensate.
\par In Fig. \ref{Fig:F6} we picture this process of merging two condensates for a Hubbard chain. The actual lattice structure is not important and such a process could be performed by driving condensates which are initialised on two halves of any bi-partite lattice. Moreover, this process could also be used to merge two condensates of the same type, such as states of the form $(S^{+})^{M/4}\ket{\chi_{1}} \otimes (S^{+})^{M/4}\ket{\chi_{1}}$ or $(\eta^{+})^{M/4}\ket{\chi_{2}} \otimes (\eta^{+})^{M/4}\ket{\chi_{2}}$. In this case the driving only needs to preserve the relevant SU(2) symmetry in order to phase-lock and merge the two condensates into a larger one.

\subsection{Experimental Implementation of SU(2) Symmetry Preservation} Finally, it is important to discuss how driving which preserves the SU(2) symmetries of the Hubbard model can be achieved experimentally. If we consider an optical lattice implementation of the Hubbard model \cite{HubbardOpticalLattice1, HubbardOpticalLattice2, HubbardOpticalLattice3} we can take advantage of the fact the Hubbard interaction and hopping strengths have a well defined relationship with the depth and separation of the potential minima which form the optical lattice sites. These quantities can be directly controlled, and made to oscillate, by modulating the standing-wave interference pattern which generates the potential landscape - a process which has already led to the experimental realisation of a Hubbard Hamiltonian with time-dependent parameters \cite{ShakenHubbard} and could be used to realise the unique states we have observed here, including the exotic $\eta$-spin condensate in Fig. \ref{Fig:F6}.
\par Moreover, in a quantum materials setting, a recent experiment has shown how laser excitation of the vibrational modes of the organic charge transfer salt ${\rm \kappa - (BEDT-TTF)_{2}Cu[N(CN)_{2}]Br}$, whose conducting layers can be described by a triangular Hubbard model, leads to the formation of transient superconducting features \cite{KappaSaltExperiment}. Density functional theory modelling of the material has shown that the laser excitation induces a periodic time-dependence in the parameters of the triangular Hubbard Hamiltonian which, in combination with the irregular geometry of the system, leads to the system transiently establishing particle-hole order as it absorbs energy from the driving field \cite{TindallFrustration}. The numerical calculations in Ref. \cite{TindallFrustration} thus provide a possible connection between the mechanism of heating-induced order and this recent experiment.
\par It is also worth mentioning that, in this work, we have considered the situation where the relevant SU(2) symmetries are completely preserved and so the correlated states we observe will form in the long-time limit and persist indefinitely - i.e they are not prethermal states but are exact steady states of the system. In a realistic experimental setup, however, these SU(2) symmetries will never be perfectly preserved due to the presence of thermal effects and lattice imperfections. In this case, assuming these unwanted mechanisms are sufficiently small in magnitude compared to the driving strength, then we expect the correlated states to instead form transiently and be observable on some intermediate timescale, prior to the system eventually heating up to a featureless infinite temperature states.

\section{Conclusion}
In this paper we have simultaneously diagonalised the dual ${\rm SU}(2)$ symmetries of the Hubbard model on an arbitrary graph. This diagonalisation has allowed us to construct the long-time states of the driven model and classify and predict their properties under various symmetry classes of driving. The preservation of either, or both, of the ${\rm SU}(2)$ symmetries leads to a significant dynamical re-ordering of the long-range correlations of the system, resulting in states with off-diagonal long-range correlations in the corresponding symmetry sectors. We have analysed how these correlations scale with the relevant initial state properties, lattice filling and the graph size. This analysis led us to identify a mechanism by which a unique condensate in the thermodynamic limit - hosting both spin-exchange and particle-hole order simultaneously - can be formed. 
\par Here, we have focused directly on the case of periodic driving, where a many-body system will generically undergo the desired heating for most non-trivial driving terms and experimental implementations are possible with current technologies. We emphasize, however, that this mechanism of 'heating-induced order' can occur in any quantum system which continuously absorbs energy from an external source whilst the requisite symmetries are preserved. For example, coupling the system to an energetic Markovian external source which introduces decoherence through local, hermitian jump operators has been shown to cause the desired heating \cite{HeatingInducedOrder, TindallSynchronisation} and driving in the form of strong kicking or non-monochromatic pulses \cite{NonMonochromatic1, NonMonochromatic2, TindallFrustration} are likely to also provide a route to inducing these correlated states. 
\par Alongside this, our work opens up several further questions. Firstly, we have not quantified the effect of the \textit{polygon} symmetries of the Hubbard lattice on the long-time states reached under periodic driving. Whilst there is a minimal effect on the geometries we consider, more complicated structures could significantly alter the pairing landscape of the system, revealing exotic states which have preferential directions for the flow of supercurrents within the lattice structure.
\par Furthermore we emphasize that this intuition of symmetry-constrained relaxation to ordered states is not limited to the single-band Hubbard model. For example, the $n$ species Hubbard model is SU(n) symmetric \cite{SU3Hubb} and driving terms which preserve this symmetry would lead to a unique state which possesses off-diagonal long-range spin-exchange correlations between each distinct pair of fermionic species. Moreover, a number of other systems, such as the SU(n) Heisenberg model \cite{SUNHeisenberg} or multi-band/ multi-orbital Hubbard models \cite{MultiBandHubbardSymmetries, MultiOrbital1}, possess special unitary/ orthogonal symmetries which could be exploited to realise similar exotic, correlated states under heating.

\begin{acknowledgments}
We would like to thank Martin Claassen, Yao Wang, Andrea Cavalleri, Michelle Buzzi and Daniele Nicoletti for helpful comments. This work has been supported by EPSRC grants No. EP/P009565/1 and EP/K038311/1 and is partially funded by the European Research Council under the European Union’s
Seventh Framework Programme (FP7/2007-2013)/ERC Grant Agreement No. 319286 Q-MAC. JT is also supported by funding from Simon Harrison. MAS acknowledges support by the DFG through the Emmy Noether programme (SE 2558/2-1) and F. S. acknowledges support from the Cluster of Excellence `Advanced Imaging of Matter' of the Deutsche Forschungsgemeinschaft (DFG) - EXC 2056 - project ID 390715994.
\end{acknowledgments}

\providecommand{\noopsort}[1]{}\providecommand{\singleletter}[1]{#1}%

\end{document}